\documentclass[usegraphicx,usenatbib]{mn2e}
\usepackage{times, amssymb}
\usepackage{epstopdf}

\title[Characterisation of M giant variability with {\em Kepler}]
{Variability of M giant stars based on {\em Kepler} photometry: general characteristics}
\author[B\'anyai et al.]{E. B\'anyai$^{1}$, L. L. Kiss$^{1,2,3}$, T. R. Bedding$^{2,6}$, B. Bellamy$^{2}$, J. M. Benk\H{o}$^{1}$, A. B\'odi$^{4}$,  \newauthor J. R. Callingham$^{2}$, 
 D. Compton$^{2}$, I. Cs\'anyi$^{4}$, A. Derekas$^{1,2}$, J. Dorval$^{2}$, D. Huber$^{2,5}$,
\newauthor O. Shrier$^2$, A. E. Simon$^{1,3}$, D. Stello$^{2,6}$, Gy. M. Szab\'o$^{1,3,4}$, R. Szab\'o$^{1}$, K. Szatm\'ary$^{4}$ 
\\
$^1$Konkoly Observatory, Research Centre for Astronomy and Earth Sciences, Hungarian Academy of Sciences, H-1121 Budapest,\\ Konkoly Thege M. \'ut 15-17, Hungary\\
$^2$Sydney Institute for Astronomy (SIfA), School of Physics, The University of Sydney,
NSW 2006, Australia\\
$^{3}$ELTE Gothard-Lend\"ulet Research Group, H-9700 Szombathely, Szent Imre herceg \'ut 112, Hungary\\
$^{4}$Department of Experimental Physics and Astronomical Observatory, University of Szeged, H-6720 Szeged, D\'om t\'er 9., Hungary\\
$^{5}$NASA Ames Research Center, Moffett Field, CA 94035, USA\\
$^{6}$Stellar Astrophysics Centre, Department of Physics and Astronomy, Aarhus University, 8000 Aarhus C, Denmark\\
}

\begin{document}

\date{Accepted 2013 September 3.  Received 2013 September 3; in original form 2013 August 5}

\maketitle

\begin{abstract}

M giants are among the longest-period pulsating stars which is why their studies were traditionally
restricted to analyses of low-precision visual observations, and more recently, accurate ground-based data.  Here we present 
an overview of M giant variability on a wide range of time-scales (hours to years), based on analysis of thirteen quarters 
of {\em Kepler} long-cadence observations (one point per every 29.4 minutes), with a total time-span of over 1000 days. About two-thirds of the sample stars have been selected from the ASAS-North survey of the {\em Kepler} field, with the rest supplemented from a randomly chosen M giant control sample.

We first describe the correction of the light curves from different quarters, which was found to be essential. We use Fourier analysis to calculate multiple frequencies for all stars in the sample. Over 50 stars show a relatively strong signal with a period equal to the Kepler-year and a characteristic phase dependence across the whole field-of-view.  We interpret this as a so far unidentified systematic effect in the {\em Kepler} data. We discuss the presence of regular patterns in the distribution of multiple periodicities and amplitudes. In the period-amplitude plane we find that it is possible to distinguish between solar-like oscillations and larger amplitude pulsations which are characteristic for Mira/SR stars. This may indicate the region of the transition between two types of oscillations as we move upward along the giant branch.

\end{abstract}


\section{Introduction}

M giants are long-period variables requiring years of continuous observations for their study. Much of our recent knowledge was gained from microlensing surveys of the Magellanic Cloud and the Galactic Bulge, such as MACHO \citep{Wood1999,Alard2001,Derekas2006,Fraser2008,Riebel2010}, 
OGLE \citep{KissBedding2003, KissBedding2004,Ita2004,Soszynski2004,Soszynski2005,Soszynski2007,Soszynski2009,Soszynski2011,Soszynski2013} and EROS \citep[]{Lebzelter2002,Wisniewski2011,Spano2011}.

While analysing photometric data of red giants in the MACHO survey of the LMC, \citet{Wood1999} found several sequences in the period-luminosity (P-L) plane, which were labelled as A, B, C, E and D, representing shorter to longer periods. Subsequent studies have shown that these structures of the sequences is rich, with over a dozen features that are related to luminosity (below or above the tip of the Red Giant Branch - see, e.g., \citealt{KissBedding2003,Fraser2008}) and chemical composition \citep[carbon-rich vs. oxygen-rich, ][]{Soszynski2009} or might have a dependency on the wavelength range of the luminosity indicator \citep{Riebel2010}.  
The most distinct parallel sequences A and B represent the radial overtone modes of semiregulars (SR). These stars are numerous and most of them have multiple periods. The Miras lie on sequence C, which corresponds to the fundamental mode \citep{Wood1999,XiongDeng2007,Takayama2013}. The power spectra of semiregulars that are observed for a large number of pulsation periods show modes with solar-like Lorentzian envelopes \citep{Bedding2003,Bedding2005}. This suggests that stochastic excitation and damping take place. With decreasing luminosities the pulsations decrease in amplitude and become more difficult to detect. However, these also have shorter periods, making them good candidates for high-quality space photometry from {\em CoRoT} and {\em Kepler}. In addition to the mentioned sequences of the SRs and Miras, there are two sequences in the P-L plane: sequence E and D representing the eclipsing binaries and  the Long Secondary Periods, respectively \citep{Wood1999}. The latter remains unexplained \citep{Nicholls2009,WoodNicholls2009,Nie2010}.
 
Although there is significant improvement in the understanding of M giant variability, there remain many question regarding the excitation and damping of the pulsations, and the expected crossover from Mira-like to solar-like excitation \citep{Dziembowski2001} which must take place in M giants. While the period-luminosity relations seem to be universal, regardless of the galactic environment \citep[see ][]{Tabur2010}, the full potential of these stars as tracers of the galactic structure is yet to be fully explored. The presence of many modes of oscillation is expected to enable the application of asteroseismology for the most luminous giants \citep{Dziembowski2012,DziembowskiSoszynski2010}, which may be affected by the mass-loss in the upper parts of the giant branch. There has also been some controversy on the short-period microvariability of Mira-like stars \citep{deLaverny1998,Wozniak2004,Lebzelter2011}, and {\em Kepler}  might prove to be ideal for resolving this issue. The complex light curves have also been interpreted in terms of stochasticity and chaos \citep{KissSzatmary2002,Buchler2004,Bedding2005}.

Working Group 12 (hereafter WG12) of the Kepler Asteroseismic Consortium \citep{Gilliland2010} was formed for the purpose of studying Mira and Semiregular pulsations in the {\em Kepler} data. Here we present the first results obtained from the analysis of the WG12 sample. The paper is organised as follows.  Sec. \ref{section:WG12Sample} presents a detailed description of the WG12 stars, which indclude the selection criteria. Sec. \ref{section:dataAnalysis} describes the data analysis. Sec. \ref{section:discussion} presents the comparison of our results with ground-based photometry, the study of frequencies and amplitudes of light-curve variations and time-frequency analyses. A brief summary is given in Sect. \ref{section:summary}.

\section{The WG12 sample and its {\em Kepler} observations}\label{section:WG12Sample}

\begin{figure}
\begin{center}
\includegraphics[angle=270,width=8cm]{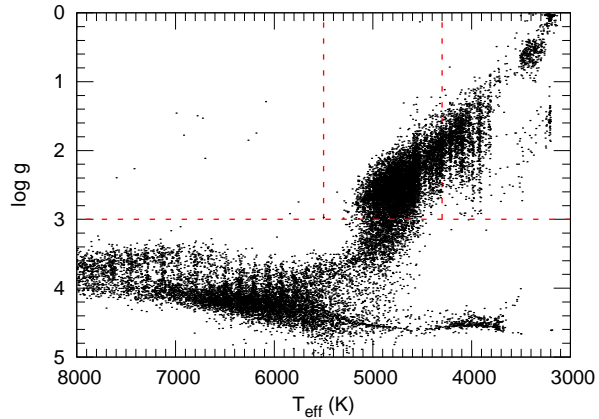}
\caption{Surface gravity vs. effective temperature from the Kepler Input Catalogue (KIC). All stars with KIC magnitude $<12$ are plotted. The bright K giants are confined between the two red vertical dashed lines, while the M giants were selected from the upper right region.}
\label{image:WG12targets}
\end{center}
\end{figure}

M giants are the longest period variable stars in the KASC program. The typical time-scale of variability is of the order of one or two Kepler quarters, which means removing instrumental drifts is potentially difficult. On the other hand, their amplitudes are above the usual instrumental effects, so correcting the M giant light curves should be a relatively simple task (and involves essentially neglecting every systematic effect that
goes beyond a constant vertical shift in the light curves from quarter to quarter). After having combined three years of {\em Kepler} data (we used the quarters Q0 $-$ Q12), we can characterize M giant variability in a homogeneous and meaningful way. 

The total sample includes over 300 M giant stars. We had two lists of targets: one was initially selected from a dedicated northern ASAS3 variability survey of the Kepler field \citep{Pigulski2009}, while the second one was selected from the Kepler Input Catalogue \citep{Brown2011}. The first set of M giant targets have been selected by combining $T_{\rm eff}$ and $\log g$ values from KIC, the J$-$K colour from 2MASS and the variability information from the ASAS3 survey. We adopted $T_{\rm eff} < 4300$ K, $\log g < 3.0$ and restricted the sample to Kepler magnitude $<12$ (see in Fig. \ref{image:WG12targets}). A cross-correlation with ASAS3 resulted in 317 stars, which were further cleaned by removing problematic cases (e.g., crowding index $< 0.95$ or an ASAS3 variability that is incompatible with a red giant). That resulted in 200 targets with variability information. Since the ASAS3 variables are all cool and large amplitude stars, we created a second list of further 200 target candidates with warmer M giants randomly selected from KIC10 with the same limits ($T_{\rm eff} < 4300$ K and $\log g < 3.0$). These stars are expected to show small-amplitude pulsations that were not detectable with the ASAS3 survey. The final list of targets that was approved for observations by {\em Kepler} contained 198 stars from the ASAS3 variable list and 119 from the other list. Most of these 317 stars have uninterrupted long-cadence (one point per every 29.4 points and short gaps between the quarters)  coverage throughout Q0 to Q12 and their data were analysed using the process described below.

\section{Data analysis}
\label{section:dataAnalysis}
\subsection{Correcting M giant light curves}

\begin{figure*}
\includegraphics[width=\textwidth]{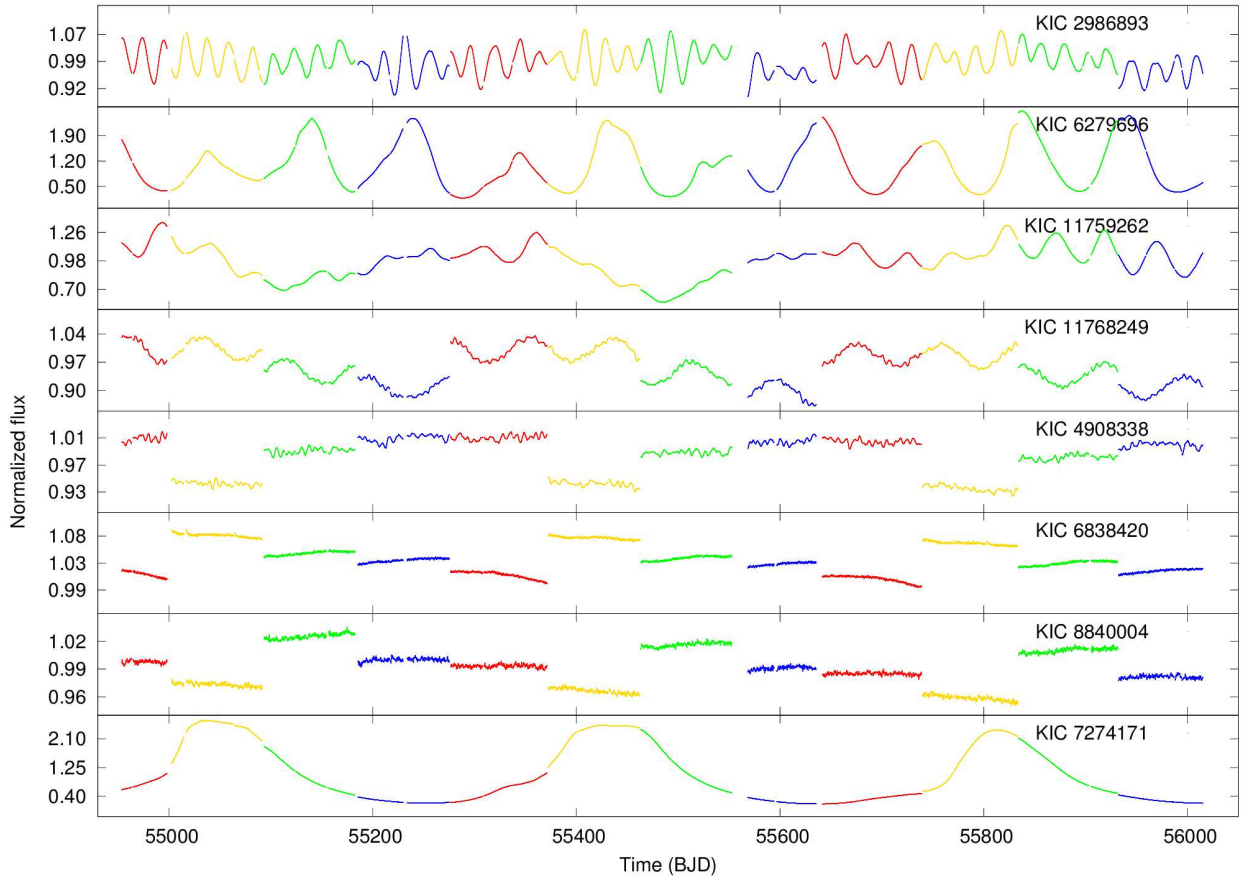}
\caption{Light curves of various M giants emphasising flux jumps of stars with different light variations. The x-axis represents time in barycentric Julian days (BJD). (See the electronic version of the article for the figure in colours.)}
\label{image:fluxJump}
\end{figure*}

The {\em Kepler} space telescope rolls 90 degrees every quarter of a year, and consequently,  variability of the majority of target stars is measured by a different CCD camera and using a slightly different aperture every quarter in a cycle of a year. For M giants with variability time-scales comparable to a yearly quarter, there is great difficulty distinguish quarter-to-quarter variations from the intrinsic stellar variability \citep{Gilliland2011}. 
\begin{figure}
\includegraphics[width=8cm]{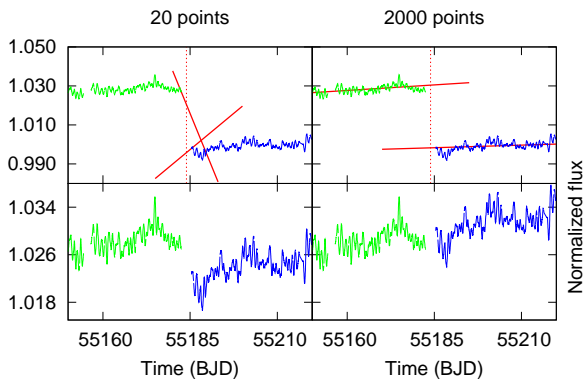}
\caption{An example for the dependence on the number of points selected for the linear fits when correcting for the quarter-to-quarter jump.}
\label{image:stitching}
\end{figure}

\begin{figure*}
\includegraphics[width=\textwidth]{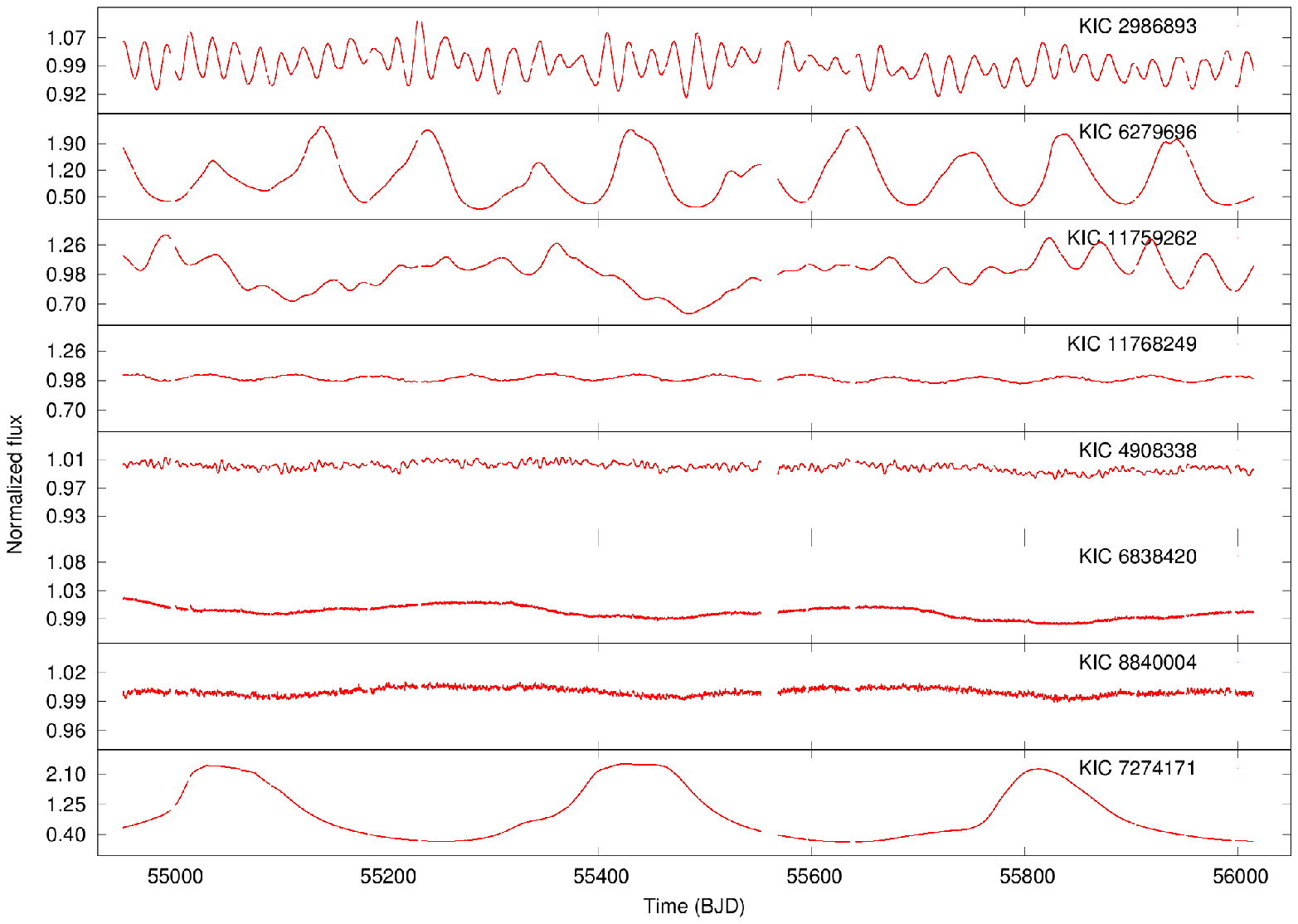}
\caption{Data for the same stars as in Fig.\ 2 after the correcting procedure.}
\label{image:stitchedLC}
\end{figure*}

Fig. \ref{image:fluxJump} presents raw light curves of eight stars. Clearly, some of the light curves (e.g. KIC 6279696, KIC 7274171) have smaller jumps and are more smoothly connected than others. Given the wide range of frequency and amplitude of the variability of the targets, it is not possible to  use a single method with the same parameters for correcting all stars.

\citet{Garcia2011} discussed in detail how the majority of KASC targets (solar-like oscillators, lower luminosity red giants) have been corrected for outliers, jumps and drifts in the data that have several different causes. Most importantly, the rotation of the telescope introduces a quasi-regular cycle of systematic jumps in the mean flux, reflecting the fact that the pixel mask used for photometry does not capture all the flux. 
Pixels with low signal were deliberately discarded, which was to optimize transit detection but made absolute photometry impossible on long time scales. The procedures described by \citet{Garcia2011} work well for the rapidly oscillating stars, with variability on time scales shorter than 10 days.

\citet{Kinemuchi2012} also discussed several general methods for correcting the flux jumps between the quarters. One proposed method is to align the time-invariant approximations for crowding and aperture flux losses. A disadvantage of this method is that the correction factors are model-dependent and are averaged over time, whereas in practice they do vary with time. Another possibility is to normalize each light curve by a functional fit or a statistical measure of the data. However, this method might introduce non-physical biases into the data. The third method is to increase the number of pixels within the target mask, although this will introduce additional shot noise into the resulting light curve.

The above-mentioned methods are best-suited for rapid variables or transiting exoplanet systems, where a smooth averaging (high-pass filter) does not distort the stellar signal. For an M giant where the length of the quarter is comparable to the mean time-scale of variability, we should be cautions about methods that were developed for other types of stars. Because of that, we decided to follow a simple procedure to correct only for the flux jumps, leaving all the low-frequency signals (both stellar and instrumental) untouched. 
Linear fitting and extrapolation were used to remove the quarter-to-quarter offsets.
For each jump we fitted lines to a number of points before and after the jump. Next, we extrapolated both of the lines to the center of the gap. 
The difference between the two lines represents the amount of correction required for a smooth transition between the two adjacent quarters. The flux data in the latter quarter was multiplied by this factor. 

We have developed a graphical user interface (GUI), which allows the user 
to set the fitting parameters for the program calculating the shifts. The fitting parameters include the number of points to be used for the linear fits, with the option of setting this for all quarters or  quarter by quarter. For slowly varying light curves 20 points were used for the fit, whereas for light curves that were dominated by high frequency variation, 2000 points were used (in the latter case, the linear fit averaged out the rapid fluctuations but retained the information of the slow trends). For some light curves different quarters required different sets of fitting parameters, predominantly when a quarter was missing. We applied this procedure to 317 stars of our sample, visually inspecting each light curve and adjusting the fitting parameters for each case. Fig. \ref{image:stitching} shows an illustration of why identical fitting parameters were not used: while 20 points were usually enough for getting a smooth transition between two quarters, stars with rapid fluctuations must be treated differently.

Figure \ref{image:stitchedLC} presents the light curves of the same eight stars as in Fig. \ref{image:fluxJump}. after correcting for the jumps. The program also divided the data by the mean flux level. In the case of KIC 4908338,  KIC 6838420 and KIC 8840004 we used 2000 points for the fits, 20 points were used for KIC 2986893 (except for the jump around BJD 55500)\footnote{In several cases the jump after the 7th quarter needed more points for the fit, because it was caused by a safe mode event and lasted 16 days as opposed to the ordinary 2 days.}, KIC 6279696, KIC 7274171 and KIC 11759262, while the dataset for KIC 11768249 was corrected with linear fits to 200 points. Note that each of these light curves contains approximately 48,000 points.

All the corrected WG12 data analysed here are available for download through the electronic version of this paper.

\section{Discussion} 
\label{section:discussion}

\label{section:methods}
To characterise M giant variability with {\em Kepler}, we have performed several simple analyses. We compared the data to ground-based observations where available, then studied the amplitudes and periodicities with standard approaches. Finally, we looked into the time-dependent changes of the periods and amplitudes using the time-frequency distributions. 

To demonstrate the potential and properties of the data, we carried out several comparisons with ground-based photometric observations such as those of the American Association of Variable Star Observers (AAVSO) or the All Sky Automated Survey (ASAS).

Amplitude and periods were determined from the Fourier transform of the time series with the program {\tt Period04} \citep{LenzBreger2005}. In order to study the time dependent phenomena, e.g. amplitude and frequency modulation, mode switching, etc. we calculated time-frequency distributions for all stars using the weighted wavelet-Z transform code 
(WWZ, \citealt{Foster1996}).
\subsection{The {\em Kepler}-year in the data}
\label{section:KeplerYear}

\begin{figure}
\begin{center}
\includegraphics[angle=0,width=8cm]{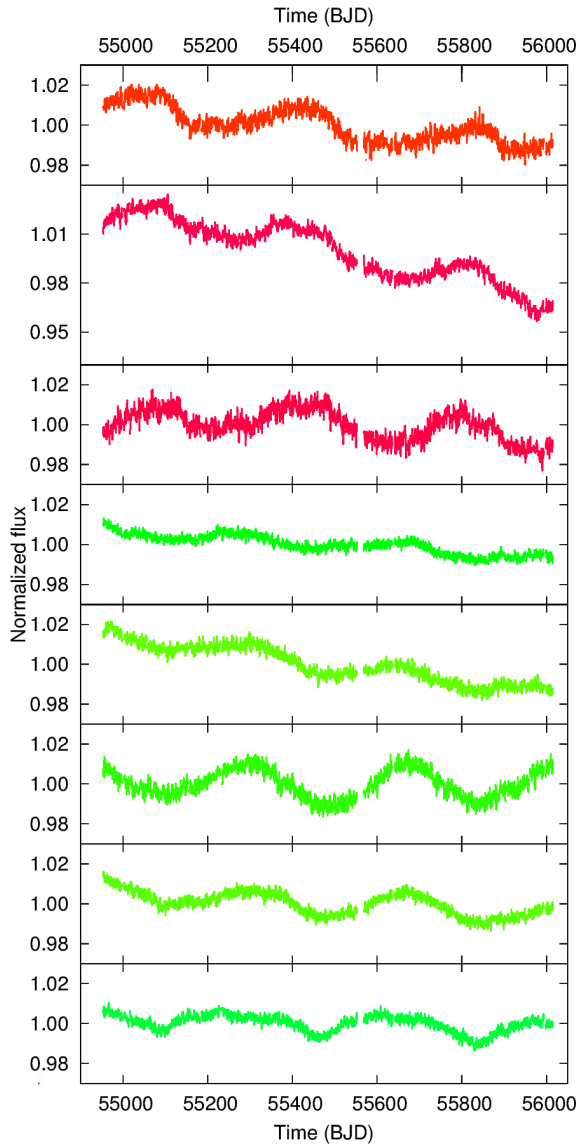}
\caption{Examples of light curves showing variablility with a period equal a {\em Kepler}-year. Colours are indicating groups of stars with different phase of the {\em Kepler-}year. Note the apparent alternation of the red (top three) and green (bottom five) light curves. (See the electronic version of the article for the figure in colours.)}
\label{image:3wave}
\end{center}
\end{figure}

\begin{figure*}
\includegraphics[angle=270, width=\textwidth]{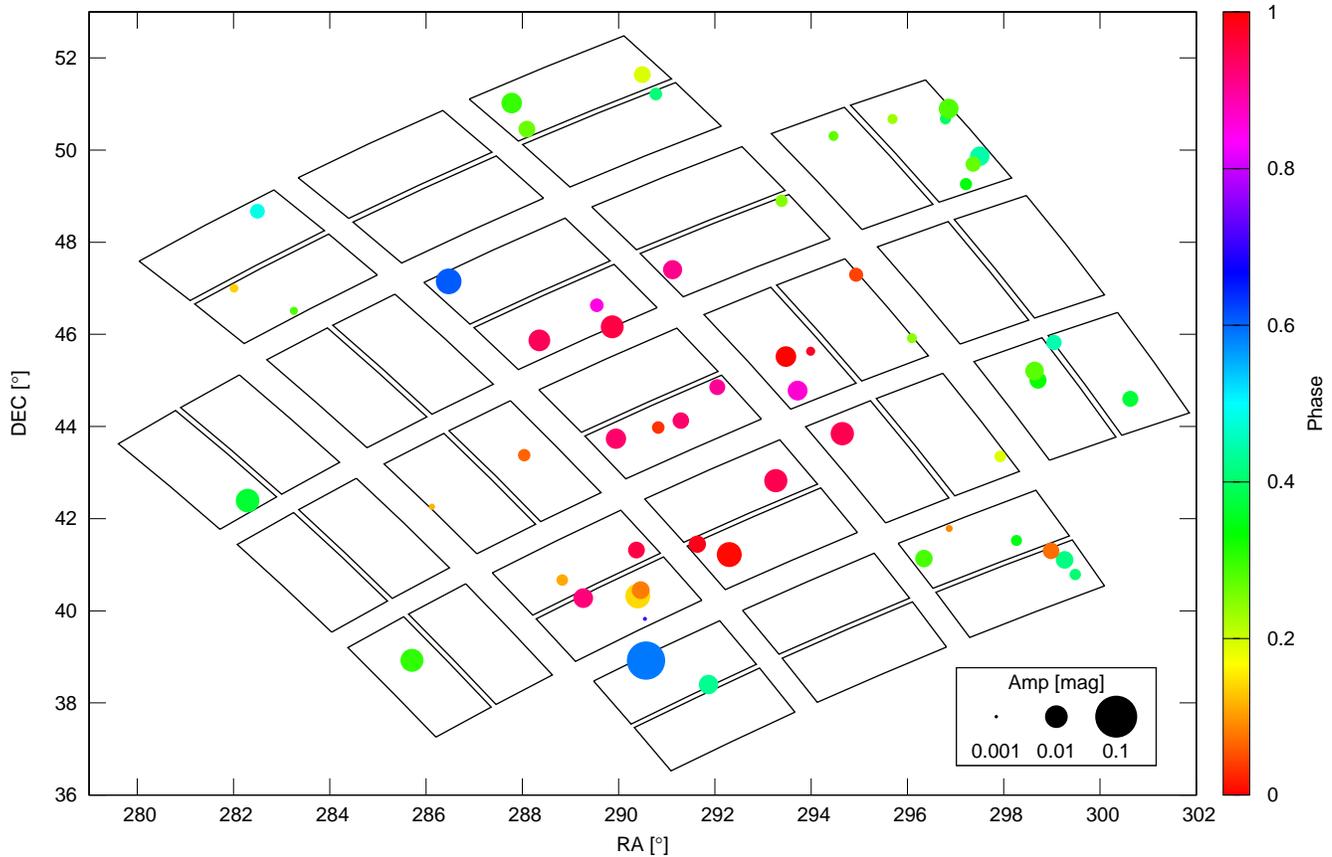}
\caption{Positions of the stars with the {\em Kepler}-year signal in the FOV. The colour bar shows the phase of the wave with a fixed zero-point. The size of the dots indicates the amplitude of the wave. (See the electronic version of the article for the figure in colours.)}
\label{image:FOV}
\end{figure*}

Visual inspection of the data revealed group of stars with similar variability. We first considered this group as 
rotationally modulated stars. However, a closer investigation of their periods and phases indicated that those changes are likely to be caused by a so-far unrecognised systematic in the {\em Kepler} data. For 56\footnote{56 out of 241 stars. If there were any missing quarters it was not possible to securely determine the presence of the {\em Kepler-year.}} stars (~23\% of the total sample) we found small variations with sinusoidal modulation
and period similar to the {\em Kepler}-year (372.5 days). This is demonstrated in Fig. \ref{image:3wave}. 
This small-amplitude fluctuation became clearly noticeable only now, after three years of data collection. 
It remains unnoticeable in the light curves of  Mira and semiregular stars due to their large amplitudes. 

To our knowledge, no {\em Kepler} Data Release Notes mention this periodicity as an existing systematic effect in the data. This trend has probably gone unnoticed because in most science investigations using {\em Kepler} data the light curves are de-trended (e.g. exoplanet studies, asteroseismology of solar-like stars or low-luminosity red giants). Very recently, \citet{VanEylen2013} noticed a systematic variation in the depth of the primary transit of the Hot Jupiter HAT-P-7b, which was found to be related to quarters of data and recurring yearly. The effect may be similar to what we find here, although a detailed comparison has not yet been made. For the slowly varying M giants, the possibility of an undiscovered systematic effect cannot be neglected. To investigate this effect, we analysed the amplitude and phase dependence of the {\em Kepler}-year signal across the field-of-view, as follows.

The phase of the {\em Kepler}-year variations varies among these stars. However, Fig. \ref{image:3wave} nicely illustrates that in
broad terms the light curves can be divided into two groups, in which the maxima and the minima of the flux are out of phase.
The positions of the stars in the CCD-array display a clear correlation with the phase. This is shown in Fig. \ref{image:FOV}, where the colour codes indicate the phase of the signal with a fixed zero point.  
The figure is dominated by the green and red colours, with most red dots located in the center and the green dots near the edge. 

In conclusion, one has to be careful when using {\em Kepler} data for investigating very long-term phenomena, such as M giant pulsations or stellar activity cycles, or any other study that needs homogeneous and undistorted data over hundreds of days. The  typical amplitude of the {\em Kepler}-year signal is around 1 percent, which is way above the short-term precision of the data. We are currently exploring whether this systematic effect can be removed by pixel-level photometry (B\'anyai et al., in prep.).

\subsection{Comparison with ground-based photometry}
\label{section:comparison}

\begin{figure}
\begin{center}
\includegraphics[angle=0,width=8cm]{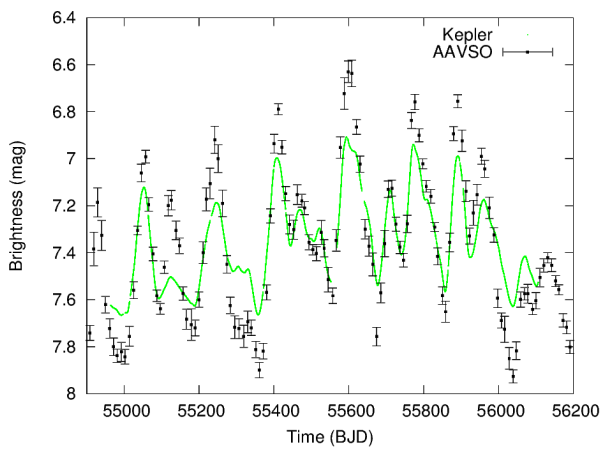}
\caption{AF Cyg AAVSO and {\em Kepler} light curves before scaling {\em Kepler} data to AAVSO.}
\label{image:AF_Cyg.v}
\end{center}
\end{figure}

We compared our {\em Kepler} measurements with ground-based photometry provided by the ASAS and AAVSO databases. 
In this section we use AAVSO data (visual, photoelectric $V$ and RGB-band  Digital single-lens reflex (DSLR) data) to study how well the ground- and space-based data can be cross calibrated. 

\begin{figure*}
\begin{center}
\includegraphics[angle=0,width=17.7cm]{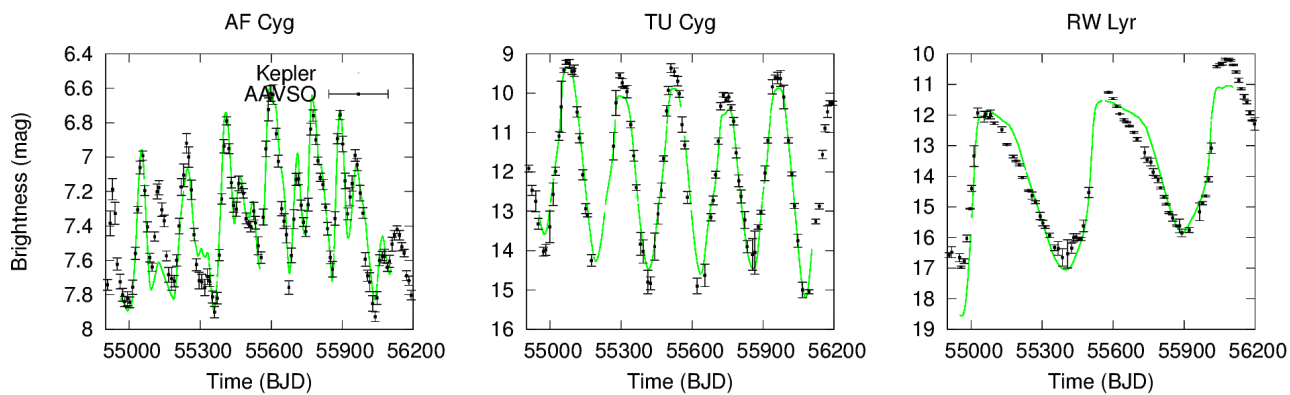}
\caption{A comparison of AAVSO and the scaled Kepler light curves for three well-known long-period variables -- AF~Cyg, TU~Cyg, RW~Lyr. In left and middle panel the AAVSO data are 10-day means of visual observations; in the right panel the brightnesses came from average measurements in Johnson V and the green channel of RGB DSLR observations. Black squares are the AAVSO data with error bars, the green dots correspond to the {\em Kepler} data.}
\label{image:AAVSOvsKepler}
\end{center}
\end{figure*}

For three well-known long-period variables we compared the AAVSO and
{\em Kepler} light curves and their frequency spectra. In Fig. \ref{image:AF_Cyg.v} we overplot ten-day means of AAVSO visual observations 
and {\em Kepler} data for the semiregular variable AF~Cyg. Here the {\em Kepler} data were converted from fluxes to magnitudes 
with zero point matched for the best fit. Although the shape of the two curves are very similar, a better fit can be achieved if the {\em Kepler} magnitudes were further scaled by a multiplicative factor. The scaled {\em Kepler} and AAVSO light curves for three stars plotted in Fig. \ref{image:AAVSOvsKepler}. The two sets of light curves are very similar even though the photometric bands were different. 

\begin{figure*}
\begin{center}
\includegraphics[angle=0, width=8cm]{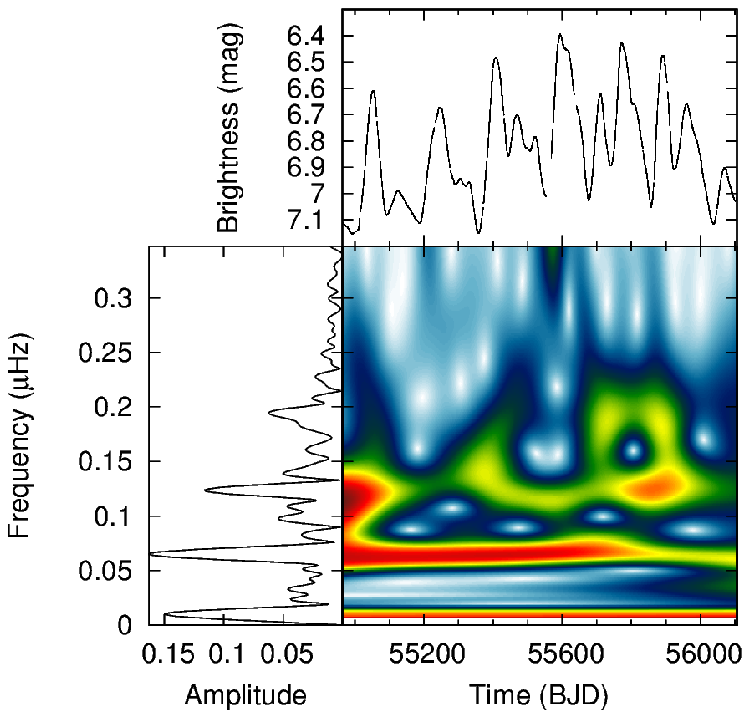}
\includegraphics[angle=0, width=8cm]{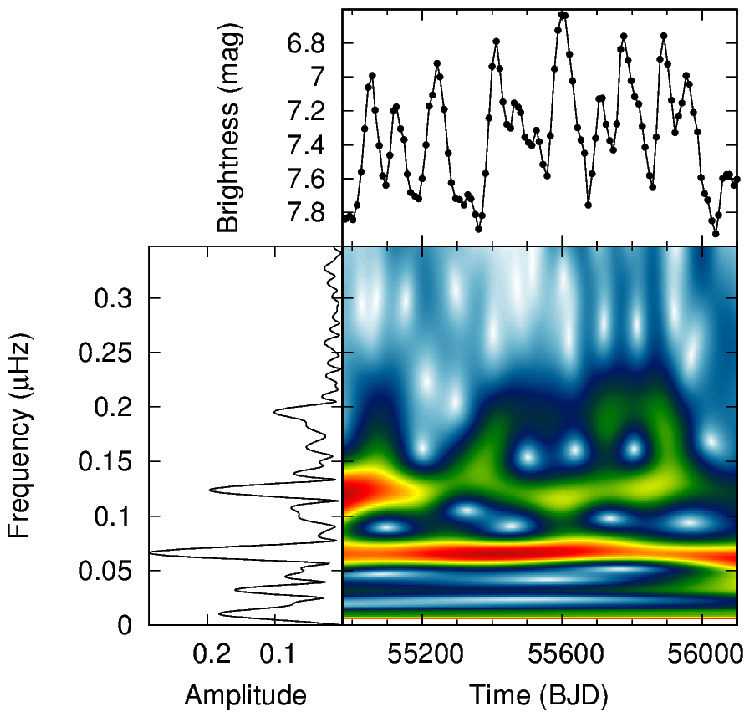}
\caption{{\it Left:} the wavelet map of AF~Cyg from the {\em Kepler} light curve. {\it Right:} the same from the AAVSO light curve.}
\label{image:AF_Cyg.wwz}
\end{center}
\end{figure*}

We noticed that in the case of RW\,Lyr a large amount of flux
has been lost due to saturation. This might be the cause of the
imperfect agreement between the {\it Kepler} and AAVSO visual data.
To remedy the situation we checked the target pixel files of this
object and verified that the assigned optimal apertures were incapable
of retaining all the flux, especially around maxima (with the approx.
500-day period maxima occurred in Q3, Q8 and Q13). Around the maximum
light as much as 50-60\% of the flux was lost.

In quarters where time intervals with no flux-loss were available we applied
the method developed in \citet{Kolenberg2011}, namely we used the 
ratio of the saturated and neighbouring column integrated flux values
to correct for the lost flux. In Q3, Q8 and Q13 however, this method
failed, because flux leaked out of the downloaded pixel area during the
whole quarter, therefore no reference values could be found. In these
cases we simply scaled the heavily saturated columns to follow the
neighbouring non-saturated columns or sum of these.

This process indeed helped to match the space- and ground-based data 
in some quarters but not in others, hence we decided not to include the 
corrected light curves in Fig. \ref{image:AAVSOvsKepler}. A more detailed pixel-based analysis of 
M giants is in progress and will be published elsewhere.

The wavelet maps of AF~Cygni 
(Fig. \ref{image:AF_Cyg.wwz}) are also very similar, 
with only minor differences in the amplitude distribution. The plots are organised in such a way that the wavelet map, in which the amplitude is colour-coded and normalized to unity, is surrounded by the light curve on the top and the corresponding Fourier spectrum on the left. This way we can see the temporal behaviour of the peaks in the spectrum and also in some cases the effects of gaps in the data. 

To characterise periodicities we performed a Fourier analysis for all the corrected light curves. With iterative pre-whitening steps we determined the first 50 frequencies with {\tt Period04}. In many cases, there were only a couple of significant peaks (like for AF~Cyg), while for the lower-amplitude stars even 50 frequencies may not include every significant peak. 

The general conclusion, based on the various comparisons to ground-based data, is that there is a good correspondence between the two data 
sources in terms of the dominant periods and the shape of the light curves for the high-amplitude long-period variables. {\em Kepler}'s superior precision allows for the determination of more periods for the lower-luminosity stars, but in cases when the frequency content is simple (like for a Mira star or a high-amplitude semiregular variable), 1,100 days of {\em Kepler} data are still too short for revealing meaningful new information. However, the uninterrupted {\em Kepler} light curves should allow for the detection of microvariability with time-scales much shorter than those of the pulsations. We note here that we found no star with flare-like events that would resemble those reported from the Hipparcos data by \citet{deLaverny1998}. 

\subsection{Multiple periodicity}

The light curves  of M giant variables can be very complex, seemingly stochastic with one or few dominant periodicities. The complexity is inversely proportional to the overall amplitude. The large-amplitude Miras are known to be single periodic variables with coherent and stable light curves. The lower amplitudes are typically associated with complicated light curve shapes that can be interpreted as a superposition of multiple pulsation modes. The {\em Kepler} WG12 sample (Sec. \ref{section:WG12Sample}) shows many features in the distribution of amplitudes and periods that were found previously 
in ground-based surveys. Here, we attempt to characterise the systematic distinction between different groups of stars.

\begin{figure}
\begin{center}
\includegraphics[angle=0]{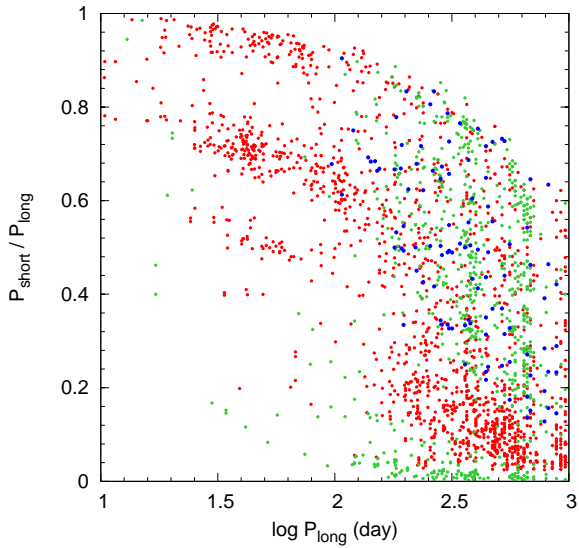}
\caption{Petersen-diagram using the first five periods for each star. Red dots refer to the ASAS group, green dots to the non-ASAS sample and blue ones to Miras. (See the electronic version for the article for the figure in colours.)}
\label{image:petersen}
\end{center}
\end{figure}

Based on the complexity in the time and the frequency domains, we sorted the stars into three groups. Stars in Group 1 have a wide range of periods between a few days and 100 days (e.g. KIC~4908338 or KIC~11759262 in Fig. \ref{image:stitchedLC}). Group 2 contains stars with very low-amplitude light curves that are mostly characterised by short-period oscillations (e.g KIC~6838420, KIC~8840004 in Fig. \ref{image:stitchedLC}), occasionally supplemented by slow changes that may be related to rotational modulation or instrumental drifts. A close inspection showed that all of them belong to the non-ASAS sample (Sec. \ref{section:WG12Sample}). Stars with light curves containing only a few periodic components (Miras and SRs) compose Group 3 (e.g. KIC~7274171 in Fig. \ref{image:stitchedLC}). In the rest of the paper we refer to these stars as Group 1, Group 2 and Group 3.

\begin{figure}
\begin{center}
\includegraphics[angle=0,width=8cm]{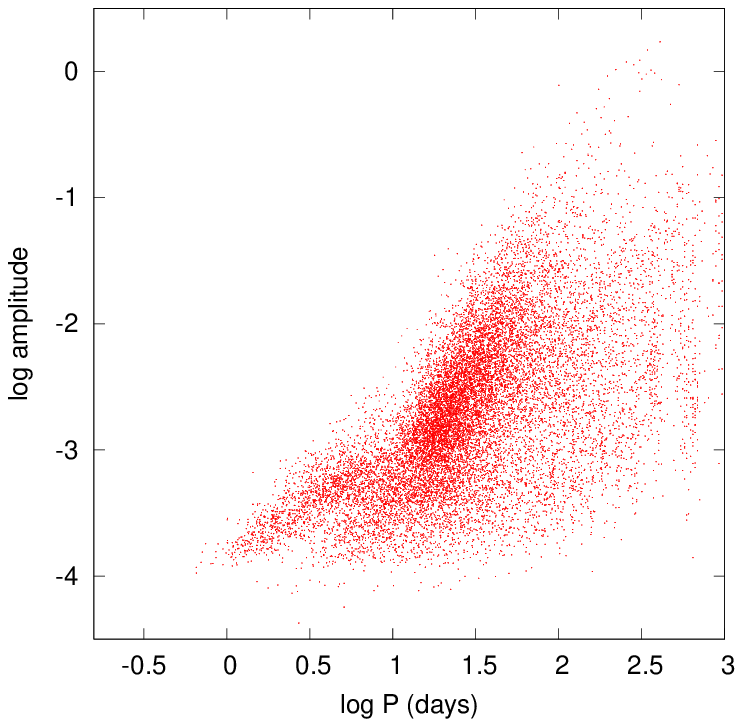}
\includegraphics[angle=0,width=8cm]{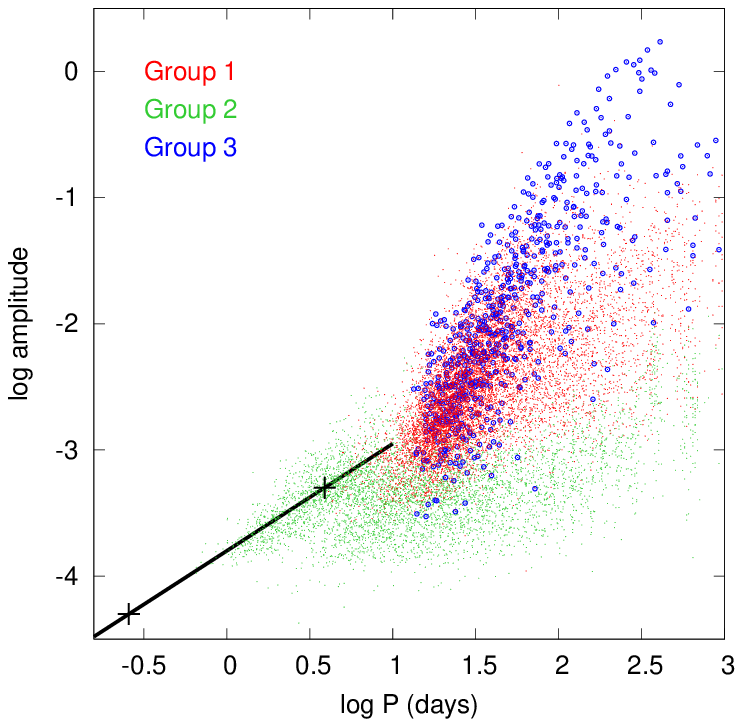}
\caption{(Top panel) the period-amplitude relations for the whole sample. (Bottom panel) different colours distinguishes the three groups. Plus signs refer to two selected points in the top panel of Fig. 3. of \citet{Huber2011} related to solar-like oscillations. The black line is drawn through these points. (See the electronic version of the article for the figure in colours.)}
\label{image:PA-relation}
\end{center}
\end{figure}

Similarly to previous studies \citep[e.g.][]{Soszynski2004,Tabur2010}, we made a Petersen diagram (period ratios vs. periods) for all WG12 stars to check any regularity in the frequency spacing and ratios. To make the diagram, we took the five most significant periods and selected every possible pair to plot their ratio as a function of the longer of the two. In Fig. \ref{image:petersen} we plotted each group with different colours. The most apparent structure is seen for periods 10-100~d, where the predominantly ASAS sample (Group 1, red dots) is clearly separated into several concentrations of points at well-defined period ratios. The most populated clump is around $P_{\rm short}/P_{\rm long} \approx 0.7-0.8$, a ratio that is known to belong to the upper Red Giant Branch stars, one to two magnitudes below the absolute magnitude of the Tip of the Red Giant Branch (TRGB, see \citealt{KissBedding2003,KissBedding2004}). The ratio is in excellent agreement with the analysis of period ratios for a similar-sized sample of bright southern pulsating red giants by \citet{Tabur2009} and those in OGLE observations by \citet{Soszynski2004}.

This ratio has been interpreted as due to pulsations in the first and second radial overtone modes in theoretical models \citep[e.g.][]{WoodSebo1996}. Recently, \citet{Takayama2013} performed detailed modelling of OGLE Small Amplitude Red Giants (OSARGs) from the OGLE-III catalogue \citep{Soszynski2009}, for which they found that the rich structure in the Petersen diagram can be explained by radial overtone modes, as well as by non-radial dipole and quadropole modes. Our plot in Fig. \ref{image:petersen} lacks the details of the sub-ridges that are so easily visible in the OGLE-III data - partly because of the lower number of stars, partly because of the significantly shorter time-span of the {\em Kepler} observations.

Another distinct clump is visible around period ratios of 0.5. This could either be related to pulsation in the fundamental and first radial overtones \citep{Takayama2013} or to non-sinusoidal light curve shapes, for which the integer harmonics of the dominant frequency appear with large amplitudes (such as the ellipsoidal binary red giants or the long-period eclipsing binaries).

Looking at the location of the green points in Fig. \ref{image:petersen}, it may seem surprising that stars in Group 2, i.e. objects characterised by their rapid variability, hardly appear in the lower left corner of the plot, where the short-period variables should fall. Instead, the green points are scattered in the long-period end of Fig. \ref{image:petersen}, which means that their Fourier spectra are heavily contaminated by low-frequency noise, so that extracting only the first five peaks in the spectra is not enough to measure the frequencies of the rapid variations. This behaviour is well documented for lower-luminosity red giants, where the regular frequency pattern of solar-like oscillations appear on top of the characteristic granulation noise in the power spectra \citep{Mathur2012}. The lack of any structure in the green points in Fig. \ref{image:petersen} (except the vertical concentration at $\log P = 2.57$ and 2.87 resulting from to the {\em Kepler}-year variability) confirms the noise-like behaviour for Group 2.

The few blue points for Group 3 stars in Fig \ref{image:petersen} do not reveal any significant structure, although they appear close to the extensions of the distinct clumps of the red points. This is expected from the similar behaviour of larger amplitude AGB variables in the Magellanic Clouds.

\subsection{Amplitudes}

For further investigations the nature of the three Groups, we studied the amplitudes and their distributions. First we made the period-amplitude plot for the whole sample, using all the 50 frequencies and amplitudes calculated by {\tt Period04}. This is shown in the upper panel of Fig. \ref{image:PA-relation}, where the presence of two distinct populations is evident. The bulk of the giants are spread in a triangular region, starting at $\log P \approx 0.5$ and $\log {\rm amp} \approx -4$ with a very well defined upper envelope pointed to the upper right corner of the plot. To the left of this upper envelope there is a distinct feature that lies between $\log P = 0$ ; $\log {\rm amp} = -4$  and $\log P = 1$; $\log {\rm amp} = -3$ which shows strong correlation between the period and amplitude. This correlation resembles the $\nu_{\rm max}$-amplitude scaling relation that has been extensively studied with {\em Kepler} data from the main sequence to red giants \citep[e.g.][]{Huber2011}.

In the bottom panel of Fig. \ref{image:PA-relation} we show the three groups with different symbols. Apparently, Group 2 populates the low amplitude part of the diagram, covering both the short-period correlation and the long-period range below Group 1 and Group 3. To validate that the correlation is indeed in the extension of the $\nu_{\rm max}$-amplitude relation for the solar-like oscillations, we added two points, marked by the large plus signs, and a line drawn through these points and extending it up to $\log P = 1$. These points are taken from the top panel of Fig. 3 of \citet{Huber2011}, where the oscillation amplitude vs. $\nu_{\rm max}$ is shown for their entire Kepler sample. We selected the $A = 1000$ ppm and the $A = 100$ ppm amplitude levels, which have mean $\nu_{\rm max}$ values of about 3 $\mu$Hz and 45 $\mu$Hz, respectively. The $\nu_{\rm max}$ values were converted to periods in days for the comparison.
Given that our amplitudes measured by {\tt Period04} are not directly comparable to those in \citet{Huber2011}, we proceeded as follows. \citet{Huber2011} measured the oscillation amplitudes following the technique developed by \citet{Kjeldsen2008}, namely calculating the amplitudes as:
\begin{displaymath}
A=(A_{\rm raw}-A_{\rm bg})\sqrt[]{T\Delta\nu/c}
\end{displaymath}
where $A_{\rm raw}$ is the raw amplitude (e.g. measured by Period04), $A_{\rm bg}$ is the amplitude of the background due to granulation/activity, $T$ is the effective time length to convert to power density, $c$ is the effective number of modes per order, and $\Delta\nu$ is the large frequency separation. For a dataset with a length of 100 d we have $T\approx86.4$ Ms (assuming no gaps).
For $\nu_{\rm max}=3~\mu$Hz, we have $\Delta\nu\sim0.6~\mu$, $A_{\rm raw}/A_{\rm bg}\approx 2$ (see Fig. 2d in \citealt{Mosser2012}), and using $c=3.04$ (as done in \citealt{Huber2011}) gives $A_{\rm raw}/A_{\rm bg}\approx 2$. This is why the \citet{Huber2011} amplitudes were divided by 2 for the line in the bottom panel of Fig. \ref{image:PA-relation}. 

The excellent agreement between the line and the period-amplitude relation for Group 2, indicates that these stars are indeed the long-period extension of the solar-like oscillations.
It is interesting to note three issues here. First, our period determination was not aimed at measuring $\nu_{\rm max}$ at all. The fact that the blind period determination leads to a recognizable detection of the amplitude vs. $\nu_{\rm max}$ scaling indicates that the frequency range of the excited solar-like modes is quite narrow (and it is actually scaled with $\nu_{\rm max}$, see \citealt{Mosser2012}, hence any period with a significant amplitude falls close to $\nu_{\rm max}$. Second, there is a quite sharp ending of the clear correlation at $\log P \approx 1$ (which corresponds to 1.2 $\mu$Hz). This may explain that for longer periods, the break in the upper envelope may be an indication of different kind of excitation that leads to Mira-like pulsations further up along the giant branch. The transition between the two types of oscillations seems to occur at $\log P \approx 1$ in Fig. \ref{image:PA-relation}, where the green dots appear in both distinct distributions of the amplitudes. Finally, for the longer period stars (red and blue points), the upper envelope of the distribution is similar to that found for bright pulsating M giants by \citet[][- see their Fig. 15]{Tabur2010}, indicating that the {\em Kepler} amplitudes can also be compared to ground-based observations.

\subsection{Power spectra}

\begin{figure*}
\begin{center}
\includegraphics[angle=0]{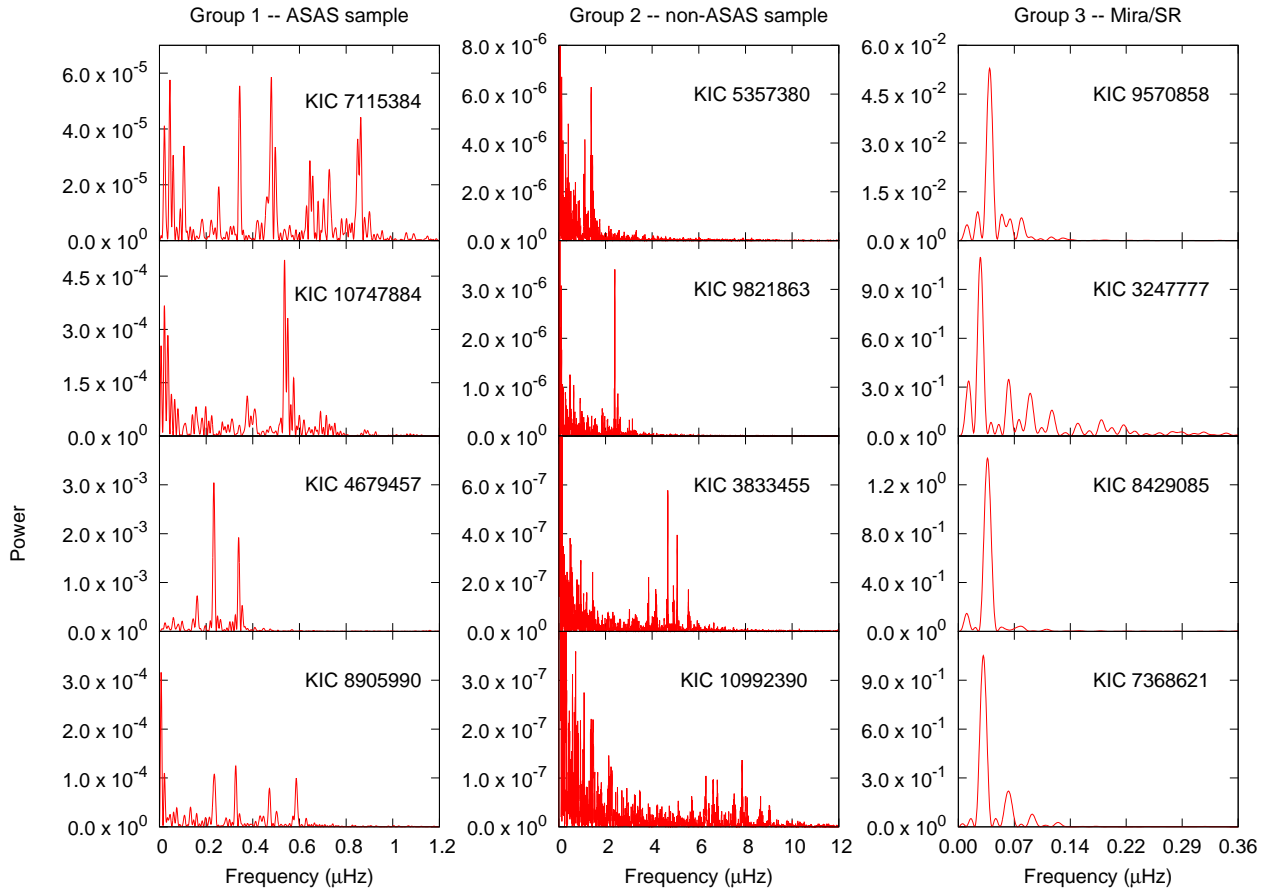}
\caption{Typical power spectra of the three Groups}
\label{image:FourierG1-G3}
\end{center}
\end{figure*}

In this Section we present typical power spectra for the three Groups of the WG12 sample. The reason for this is to illustrate the rich variety of the spectra, which indicates the complexity of M giant variability.

We show spectra for four stars in each Group in the three columns of Fig. \ref{image:FourierG1-G3}. It can be seen that in Group 1 the stars show a remarkable variety in the spectra, and most of the stars are characterized with several (up to 10-15) significant frequencies (we used the $S/N$ calculator of {\tt Period04} to quantify the significance of each peak, following the method of \citet{Breger1993}. As can be seen in the middle column of Fig. \ref{image:FourierG1-G3}, Group 2 stars are indeed typical solar-like oscillators with a granulation noise strongly rising below 2.3 $\mu Hz$ and a distinct set of significant peaks. Note that the four panels in the middle column were sorted from top to bottom by the increasing average frequency of the acoustic signal (roughly corresponding to the classical $\nu_{\rm max}$). Here every spectrum contains very high peaks (in relative sense) between 0 and 0.046 $\mu Hz$. Peaks near 0.031 $\mu Hz$ correspond to the {\em Kepler}-year and they appear in almost every spectrum with smaller or higher amplitude. Often there is another high peak near 0.008 $\mu Hz$, which is near to $1/t_{\rm obs}$, where $t_{\rm obs}$ is the length of the full dataset. The acoustic signal emerges at higher frequencies in form of some sort of structured peaks at lower amplitudes but still significant compared to the local noise in the spectrum.

Spectra for Group 3 are shown in the right column of Fig. \ref{image:FourierG1-G3}. The highest peaks appear between 0.029 $\mu Hz$ and 0.069 $\mu Hz$, i.e. periods between 150 and 400 days. In some cases, we see the integer harmonics of the dominant peak, which is caused by the strong departure from the pure sinusoidal shape, characteristic for most of the Mira stars.

\subsection{Time-frequency distributions}
\label{section:wavelet}
\begin{figure*}
\begin{center}
\includegraphics[angle=0, width=17.7cm]{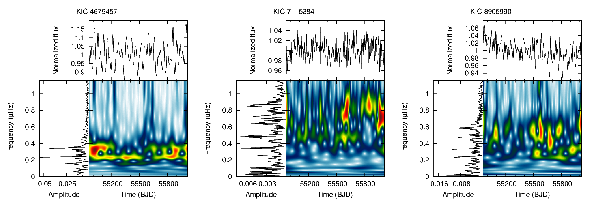}
\includegraphics[angle=0, width=17.7cm]{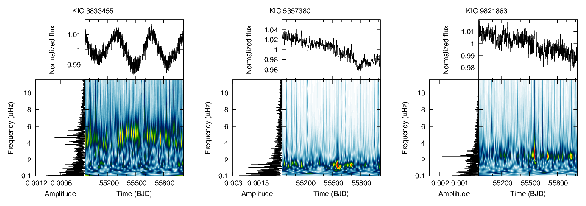}
\includegraphics[angle=0, width=17.7cm]{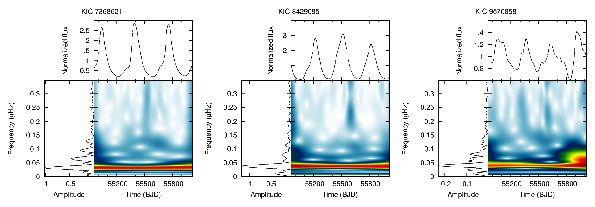}

\caption{Time-frequency distribution for selected members of Group 1, 2 and 3 (from top to bottom). Each row contains plots for three stars in each group. The most informative cases are those for Group 1, where the wavelet maps clearly indicate that many of the peaks in the Fourier spectra (left panel in each plot) corresponds to signals that have strongly time-dependent amplitudes.}
\label{image:Wavelet}
\end{center}
\end{figure*}

As a last step we surveyed systematically the time-frequency distributions for all stars in the three Groups. We searched for phenomena that are more difficult to detect with the traditional methods of time-series analysis, such as mode switching, amplitude modulation or frequency modulation. While a detailed study of all these phenomena is beyond the scope of the paper, we can demonstrate the typical cases in each Group. 

In Fig. 13 we show three examples from each Group. The three Groups were ordered in the subsequent rows from top to bottom. Group 1 stars (top row in Fig. \ref{image:Wavelet}) have multiperiodic light curves (P=10-50 days). The frequency content is rarely stable, where most of the peaks come and go on the time-scales of a few pulsation cycles. The amplitudes of the components change very strongly, and there is no apparent order in this. There are few cases when the strongest peaks are changing in sync with each other, like mode switching, but this is rare and difficult to distinguish from random amplitude changes. 

For Group 2 stars (middle row in Fig. \ref{image:Wavelet}), the wavelet maps were calculated from 0.12 $\mu Hz$ frequency (without the long-period trends) in order to get a clearer picture of the shorter and smaller amplitude variations between 1.16-5.74 $\mu Hz$, where the acoustic signal is dominant. Here the random changes of the frequencies are very apparent, a behaviour that naturally arises from the stochastic excitation of the solar-like oscillations. For Group 3 stars, the available time-span only allows measuring the stability of the dominant peak, in good agreement with the Mira character.

\section{Summary}
\label{section:summary}
In this paper we studied the global characteristics of light variations for 317 red giant stars in the {\em Kepler} database, containing 198 already known variable stars observed by the ASAS North survey and 117 stars in a control sample selected based on their estimated physical parameters. The main results of this study: 

\begin{enumerate}
\item The study of M giants with {\em Kepler} poses new challenges because of the time-scale of variability that is comparable to the length of the {\em Kepler} quarters. Most of the usual methods for correcting the trends and jumps are thus not applicable. After extensive testing we ended up with a simple light curve stitching method, which is based on linear fits at the edges of the quarters and then matching the quarter-to-quarter shifts for creating the smoothest possible light curves. We developed a software with a user-friendly GUI, which made it easy to set the fitting parameters and stitch together each quarter. When the data contain missing quarter(s), no unique solution is possible.

\item Three years of observations revealed a so far unnoticed systematic fluctuation in the data, at the levels of up to 1-3\%. We found that the period equals to one {\em Kepler}-year and the phase behaviour is clearly correlated with the position in the whole {\em Kepler} field-of-view. It is not yet clear if a more sophisticated pixel photometry would be able to remove the artefact.

\item We compared the data with various ground based photometries (visual, ASAS CCD, AAVSO DSLR, etc.) and concluded that for the large-amplitude stars, {\em Kepler} light curves can be matched very well with the ground-based data, but the amplitudes require a significant scaling by about a factor of two. {\em Kepler}'s main advantage for these slow variables is the uninterrupted observations at high-precision and high-cadence (relative to the pulsation periods).

\item We studied the distributions of periods, period ratios and amplitudes. There are several regular patterns in these distributions that can be explained by the presence of several pulsation modes, some possibly non-radial dipole or quadropole modes. We find evidence of a distinction between the solar-like oscillations and those larger amplitude pulsations characteristic for Mira/SR stars in the period-amplitude plane. This may show the transition between two types of oscillations as a function of luminosity.

\item The power spectra and wavelets reveal very complex structures and rich behaviour. Peaks in the spectra are often transient in terms of time-dependent amplitudes revealed by the wavelet maps. The overall picture is that of random variations presumably related to the stochasticity of the large convective envelope.
\end{enumerate}

With this paper we highlighted the global characteristics of M giant stars seen with {\em Kepler}. There are several possible avenues to follow in subsequent studies. Given the time-span and the cadence of the data, an interesting avenue of investigation is to perform a systematic search for rapid variability that can be a signature of mass-accreting companions. One of the archetypal types of such systems, the symbiotic binary CH Cyg, has been both KASC target and Guest Observer target, and its data can be used as a template to look for similar changes in the full {\em Kepler} red giant sample. Another possibility is to quantify the randomness of the amplitude changes using detailed statistical analysis of the time-frequency distributions. 

\section*{Acknowledgments}

This project has been supported by the Hungarian OTKA Grants K76816, K83790, K104607 and HUMAN MB08C 81013 grant of Mag Zrt., ESA PECS C98090, KTIA URKUT\_10-1-2011-0019 grant, the Lend\"ulet-2009 Young Researchers Program of the Hungarian Academy of Sciences and the European Community's Seventh Framework Programme (FP7/2007-2013) under grant agreement no. 269194 (IRSES/ASK). AD, RSz and GyMSz have been supported by the J\'anos Bolyai Research Scholarship of the Hungarian Academy of Sciences. AD was supported by the Hungarian E\"otv\"os fellowship. RSz acknowledges the University of Sydney IRCA grant. Funding for this Discovery Mission is provided by NASA's Science Mission Directorate. The Kepler Team and the Kepler Guest Observer Office are recognized for helping to make the mission and these data possible.

\end{document}